\documentclass[twocolumn,aps,amsmath,showpacs,amssymb,nofootinbib,pre]{revtex4}
\usepackage[dvips]{graphicx}
\usepackage[dvips,usenames]{color}
\usepackage{amsmath}

\begin{document}

\title{Heterogeneity effects in power grid network models}

\author{G\'eza \'Odor and B\'alint Hartmann}
\address{Centre for Energy Research of the Hungarian Academy of Sciences,
P. O. Box 49, H-1525 Budapest, Hungary}

\begin{abstract}

We have compared the phase synchronization transition of the second order 
Kuramoto model on 2D lattices and on large, synthetic power grid networks,
generated from real data. The latter are weighted, hierarchical modular
networks. Due to the inertia the synchronization transitions are of first 
order type, characterized by fast relaxation and hysteresis by varying 
the global coupling parameter $K$.
Finite size scaling analysis shows that there is no real phase transition
in the thermodynamic limit, unlike in the mean-field model.
The order parameter and its fluctuations depend on the network size
without any real singular behavior.
In case of power grids the phase synchronization breaks down at lower 
global couplings, than in case of 2D lattices of the same sizes, but 
the hysteresis is much narrower or negligible due to the low 
connectivity of the graphs.
The temporal behavior of de-synchronization avalanches after a 
sudden quench to low $K$ values, has been followed and duration 
distributions with power-law tails have been detected. 
This suggests rare region effects, caused by frozen disorder, 
resulting in heavy tailed distributions, even without a 
self organization mechanism as a consequence of a catastrophic 
drop event in the couplings.
\end{abstract}

\maketitle

\section{Introduction}

Power grids are large complex, heterogeneous dynamical system, built up
from nodes of energy suppliers and consumers. These units are 
interconnected by a network that enables energy distribution in a 
sustainable way. However, unexpected changes may cause failure that
can be described by synchronization events, which may propagate through
the whole system as an avalanche, causing blackouts of various sizes.
As the worst case these can lead to full system de-synchronization, lasting
for a long time \cite{1}. To avoid these events power grid systems 
should be designed to be resilient to local instabilities, failures 
and disturbances.
Studies have shown that valuable insights into the dynamical behavior 
of power grids can be obtained by theoretical studies that consider models 
of electrical generators, coupled in network structures, reproducing 
the topological and electrical interactions of real power grids \cite{Ace05,2}.

The so called second order Kuramoto model was proposed to describe
power grids  \cite{fila} and a number of studies exists, which focus 
on the synchronization and stability issues, such as in 
Refs.~\cite{3,4,5,12,13,18,29,32,36,Gryz}.
This is the generalization of the Kuramoto model \cite{kura}
with inertia. One of the main consequences of this inertia is that
the second order phase synchronization transition, observed in
the mean-field models, turns into a first order one \cite{TL97}.
However, according to our knowledge, the transition type, if any,
in lower dimensions has not been studied. It is well known that
discontinuous mean-field phase transitions can turn into continuous
one as the consequence of fluctuation effects \cite{odorbook}.
Fluctuation effects are enhanced in lower spatial dimensions, so
it is an open question what happens on a homogeneous, two dimensional
system. Therefore power grids may become critical, exemplified 
especially by the scale-free distributions measured on them \cite{Car}. 
This criticality has been attributed to some self-organization 
(SOC) mechanism \cite{Bak}.

On the other hand, highly heterogeneous, also called disordered 
with respect to the homogeneous, system can experience rare region
effects, that smear phase transitions \cite{Vojta}.
Rare regions, which are locally in another state than the whole,
evolve slowly and contribute to the global order parameter and
can generate various effects, depending on their relevancy.
They can change a discontinuous transition to a continuous one
\cite{round}, can generate so-called Griffiths Phases (GP)
\cite{Griffiths} or completely smear the singularity of a
critical phase transition.
In case of GP-s critical-like power-law (PL) dynamics appears over an
extended control parameter region around the critical point, 
causing slowly decaying auto-correlations and burstiness \cite{burstcikk}.
Furthermore, in the GP the susceptibility diverges with the 
system size. 
Therefore, we decided to investigate if topological
and coupling strength heterogeneities of power grids 
are strong enough to generate critical dynamics or a GP.

We generated weighted graphs of power grids, which are similar to
the real ones and large enough to allow reliable statistical physics
analysis, including finite size scaling. We created networks 
from $N\simeq 10^6$ to $N \simeq 2.3\times 10^7$ 
nodes and compared the phase synchronization transition results of 
the second order Kuramoto model with those of 2D lattices of similar sizes.

\section{Models and methods}

We have studied the second order Kuramoto model proposed by \cite{fila}
to describe network of oscillators with phase $\theta_i(t)$:
\begin{eqnarray}\label{kur2eq}
\dot{\theta_i}(t) & = & \omega_i(t) \\
\dot{\omega_i}(t) & = & \omega_{i,0} - a \dot{\theta_i}(t) 
+ \frac{K}{N_i} \sum_{j} A_{ij} \sin[ \theta_i(t)- \theta_j(t)] \ , \nonumber
\end{eqnarray} 
where $N_i$ is the number of incoming edges of node $i$,
$a$ is the damping parameter, describing the power dissipation,
$K$ is the global coupling, related to the maximum transmitted power
between nodes and $A_{ij}$, which is the weighted adjacency matrix 
of the network, containing admittance elements.

The (quenched) heterogeneity comes into the model in two ways: via
$\omega_{i,0}$-s, as intrinsic frequencies of the nodes and via $A_{ij}$, 
which describes both the topology and the admittances of the power grid.
As for the intrinsic frequencies we used uncorrelated Gaussian
random variables, with the distribution centered around the mean 
$\langle\omega_i\rangle=50$ and unit variances to model real AC system, 
although the results have been found to be invariant for this value.
For the damping parameter we assumed: $a = 1, 3$.

We have studied three different types of networks:
\begin{itemize}
\item fully connected, to recover mean-field results
\item 2D lattices, with periodic boundary conditions, simulating 
homogeneous electric power grids
\item synthetic hierarchical modular ones, generated randomly, 
following the characteristics of real electric power grids.
\end{itemize}   

\subsection{Description of the synthetic power grids}

Analysis of the electric power system often requires the use of network models to a certain extent; 
however the specific examinations largely affect the nature and the quantity of networks that are 
necessary to produce authentic results. In certain cases, it is sufficient to perform analysis 
on one or only a few networks. These usually represent either high-voltage (HV) transmission and 
sub-transmission systems or medium- and low-voltage (MV and LV) distribution systems; 
the mixed use of these networks for the same scope is rare. In case of HV networks, analysis can 
be based on network data acquired from utilities and system operators, since the volume of the 
data is limited in this case, and most of this information is also openly available. 
This is partly the reason for the over-representation of HV networks in the field of power grid 
network analysis~\cite{Pagani}.
In case of MV and LV networks however, another solution is necessary to perform extensive analysis.

One possible solution is to acquire data of so-called representative or reference network models (RNM). 
RNMs are often used tools, when future grid expansion scenarios have to be compared from the perspective 
of infrastructural needs, maintenance costs or power losses. Two common methods are used to create such RNMs. 
The first approach is based on real network data of the utilities; by applying clustering techniques 
the most typical topological configurations are identified. The literature discusses several methods 
to create RNMs, a deep and thorough review is presented by~\cite{Gomez}. The disadvantage of this 
method is that it results only a limited number actual networks, which do not provide sufficient 
variability for our examinations. The second approach is used in case no real network data is available, 
and synthetic networks are built. Widely used and known examples for such synthetic networks are 
the IEEE Bus systems, which are long-time cornerstones of network-related studies in the power 
engineering field. The necessity of synthetic networks has been highlighted by several publications 
during the last couple of years. Ref.~\cite{Wang} emphasized in their work that future power 
engineering problems are in the need of appropriate randomly generated grid networks, that 
have plausible topology and electrical parameters. They have also concluded that the admittance 
matrix has peculiar features that follow statistical trends. The Generalized Random Graph Model 
is used to generate synthetic networks by \cite{Pahwa}, but the node count of the introduced 
networks are by magnitudes smaller than it is necessary for our studies. 
Similar problems are faced with the dual-stage method of \cite{Ma}, where node count is in the 
range of thousands. For the examinations shown in present paper, the authors have developed 
a new power grid network generator algorithm, which has significant differences compared to 
the existing ones. As these differences are related to the aim of providing a realistic 
recreation of real power-gird networks, main modeling assumptions and goals are discussed 
in the following.
\begin{figure}[h]
\centering
\includegraphics[width=8cm]{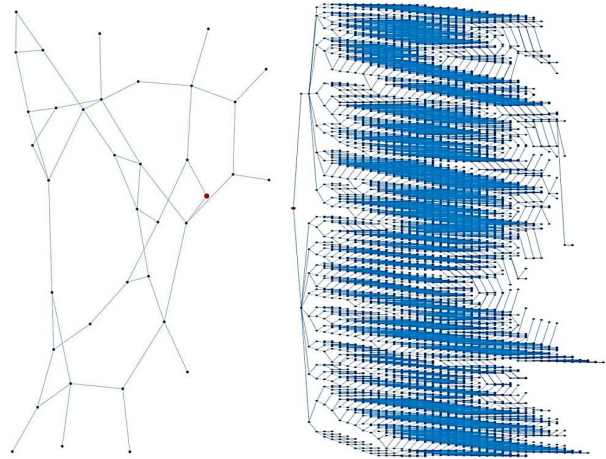}
\caption{\label{network}
Structural representation of the synthetic networks.
Left side: HV, right side: a radial cabinetwork.
The highlighted red node connects the two “layers”.
The network on the picture has $68850$ nodes and $68849$ edges.}
\end{figure}

The task of the power system is to provide cooperation between power plants, create 
interconnection on national and international level and to transmit and distribute 
the produced electricity. 
To achieve these goals at minimum ecological and economic costs, the structure of power 
systems has evolved so that transmission and distribution networks have significantly different 
characteristics. 
When designing the sample networks for current work, aim of the authors was to replicate 
functionality of real power systems, thus those two levels were handled differently. 
While admittance matrix of the transmission network is based on a real-life example 
(the Hungarian power system), matrix of the distribution network is the result of 
synthetic grid modeling.

\begin{figure}[ht]
\centering
\includegraphics[width=8cm]{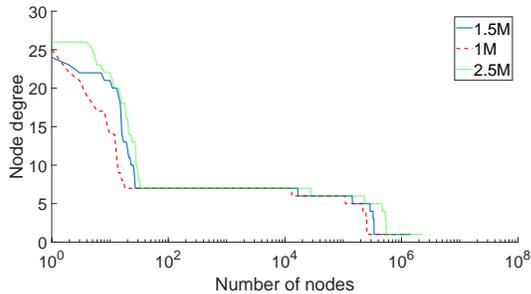}
\caption{\label{degree}
Node degree distributions of the synthetic power grids generated
for 2.5M, 1.5M and 1M networks (right to left curves).}
\end{figure}

The transmission level of a power system has to handle the largest blocks of power, 
while interconnecting major generators stations and loads of the system. 
To achieve best overall operating economy or to serve technical objectives best, 
energy flows in the transmission system can be routed, generally, in any 
desired direction. The topology of the transmission system tends to obtain a loop 
structure, which not just provides more path combinations, as no designated 
flow directions are found, but ensures an increased level of security. 
Each node of the network can receive power through multiple connections, 
thus the system is tolerant to single failures (so-called ($N-1$) criterion).

\begin{figure}[ht]
\centering
\includegraphics[width=8cm]{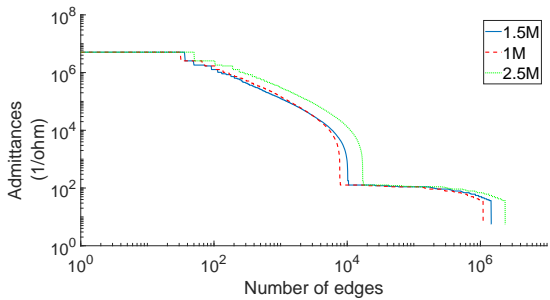}
\caption{\label{weights}
Admittance distribution of the power grids generated
for 2.5M, 1.5M and 1M networks (right to left curves).}
\end{figure}

Considering its current functionality and structure, former sub-transmission 
networks have to be handled similarly to transmission networks, although 
certain differences are to be noticed. Sub-transmission networks are usually 
designed to have a designated power flow direction from source to sink and 
have a mixed loop-radial topology.
In Hungary, the transmission network mainly consists of $750$, $400$ and $220~kV$
lines and substations, while the nominal voltage of the former sub-transmission 
level is $120~kV$. The security of delivery is increased such that both the 
$220-400~kV$ and the $120~kV$ network is meshed, and many parallel (double) 
lines are also operated. 

\begin{table}
 \caption{\label{tab:networks}
  Power-grids generated and studied.
 }
 \centering
 \begin{tabular}{|l|l|l|l|l|l|}
  \hline
  Network   & $N$       & Edge no.  & $L$                & $C^{\Delta}$ & $C^{W}$ \\
  \hline
  $1$M      & $1098583$ & $1098601$ & $1.7440\times10^6$ & $0$	      & $0$ \\	
  \hline
  $1.5$M    & $1455343$ & $1455367$ & $1.0457\times10^6$ & $0.0594$   & $0.0486$ \\
  \hline
  $2.5$M    & $2356331$ & $2356360$ & $1.6162\times10^6$ & $0.0851$   & $0.0586$ \\
  \hline
  $23$M     & $23551140$& $23551254$& $2.1129\times10^6$ & $0.0626$   & $0.0741$ \\
  \hline
 \end{tabular}
\end{table}

The distribution level of a power system constitutes the finest meshes in the overall network. 
The circuits are fed from sub-transmission level ($120~kV$) and supply electricity to the small 
(residential) and medium-sized (small industrial and commercial) customers. 
The topology of this network is dominantly radial, thus nodes have fewer connections 
compared to the transmission networks. The primary distribution level ($20$ and $104~kV$) is 
fed directly from the $120~kV/MV$ substations. The MV feeders cover wider supply areas 
and each feeder supplies multiple distribution transformers. These transformers provide 
connection between the primary and the secondary distribution level. 
The latter on is operated at $0.4~kV$ nominal voltage. 

Due to the functional and topological characteristics, the node number of distribution 
networks is by magnitudes bigger than as of transmission networks. On one hand this 
characteristic makes distribution grids a suitable choice for the examination of 
synchronization transition of networks. On the other hand, examination of real 
topologies would require a large collection of electrical and topological data, 
which is usually not openly available from utility companies, thus synthetic grid 
modeling is favored to recreate this part of the power system.

As it was shown previously, a number or publications discuss the possibilities of 
both clustering power grids and creating synthetic topologies for analysis. 
One of the common weaknesses of these methods is that they dominantly focus on 
HV and MV networks, which have limited number of nodes, insufficient for our studies. 
To present a rough comparison, the proportion of the number of HV, MV and LV 
nodes in a power system is in the range of 1:100:10000, respectively. 
The only field, where LV networks are extensively studied, is the area of reference 
networks models, which are used to determine power losses of the network, 
but in this case usually only a set of representative networks are created, 
which is limiting the number of topologies to be examined. 
In contrast for present paper the authors have generated random power system 
topologies consisting of a few million nodes. The other main difference between 
the processed literature and our method is that the present work uses solely weighted 
graphs, while the cited ones rely mostly on unweighted ones, which ignore 
valuable information on the behavior of the power system. 
Another significant extension of the authors' model is that transformers are 
represented as weighted bi-node connections, instead of the typical choice 
of handling the two terminals of the transformer as a single node. 
With this extension the node and connection number of the admittance matrix is increased 
and the node degree distribution is also affected.

To generate the random topologies, the authors have used an iterative process in MATLAB. 
The initial step of the process it to set up the transmission and sub-transmission 
levels (lines and transformers) and to mark all $120~kV$ substations. 
In the second step a random number is generated to determine the nature of the 
connected MV network; in Hungary approximately one-third of all MV networks are 
cable lines (operated on $10~kV$) and two-third are overhead lines (operated on $20~kV$). 
It is important to distinguish these voltage levels not only because of different 
admittance values but also because of their different topological characteristics 
(line length, transformer nominal power, number of feeders, etc.). 
After the voltage level is determined, the $120~kV/MV$ transformer is created. 
Nominal power (and thus admittance) of the unit is selected using the empirical 
distribution of such units' nominal powers. As the next step, length of the MV 
feeder main and branch lines is calculated, and the position of MV/LV transformers 
is selected along the lines. Electrical parameters of the lines are also based 
on empirical distributions and actual per length line admittances. 
As the last process of the topology generation, bi-node connections representing 
MV/LV transformers are created, and the LV radial network is generated in a 
similar way as it was shown with the MV. In the final step, individual LV consumers 
are added; this step largely increases the number of nodes with single connection 
in the network, affecting thus the node degree distribution of the graph 
representation as well.

\subsection{Analysis of the synthetic power grids}

The number of nodes in networks that are generated with the previously described 
process is approximately $N=23$ million, which is already sufficient to use for 
modeling synchronization processes, but computation times are also slowed down 
significantly. To find the golden mean of network size and computation times, 
the authors have decided to reduce these networks, while preserving its typical 
characteristics. As a result, networks with few ($1-3$) million nodes were generated, 
using the same iterative process as described before. Network analysis was performed 
on these networks, the result of which is presented in the following, using three 
example networks with approximately $N=1, 1.5$ and $2.5$ million nodes. 
To represent the structure of these networks, Fig.~\ref{network} is used an example. 
The left side of the figure shows the looped HV network, while on the right 
side the radial network of a HV node is plotted. 
It can be seen, that the structure of the radial network is similar to a tree, 
with relatively low node degrees and practically zero clustering coefficient. 

The degree distribution of the networks on Fig.~\ref{degree} shows that only a 
limited number of nodes have high degrees. This is again due to the radial 
structure of the system, where only looped sub-networks are considered central 
parts of the network. The high number of nodes with $k=5$ and $k=6$ degrees
represent LV feeders, where 3 or 4 end-users are connected to the same 
nodes of a radial network. The admittance distribution on Fig.~\ref{weights}
is composed of a low and a high value region, the latter exhibits a tail, which can 
be fitted linearly for $17.100 1/\Omega < Y_{ij} < 93.000 1/\Omega$.
To compare our results with those of the weightless networks we used the normalized 
admittances as weights:
\begin{equation} \label{wnorm}
A_{ij} = Y_{ij}/\langle Y_{ij} \rangle   , 
\end{equation}
by averaging over the directed edges of the networks.

Further graph measures for four example networks is shown in 
Table~\ref{tab:networks}, including the most important metrics. 
The average shortest path length is
\begin{equation}
L = \frac{1}{N (N-1)} \sum_{j\ne i} d(i,j) \ ,
\end{equation}
where $d(i,j)$ is the graph (topological) distance between vertices $i$ and $j$.
Considering the clustering coefficient, as it was 
shown previously, as vast majority of the network (including more than 99.995\% 
of the nodes) has a tree structure, the value of the coefficient is near zero
and the small differences are caused by the structure of the central looped 
sub-networks. Thus clustering coefficients of these sub-networks are included 
in the table. The sub-networks consist of $37, 49, 60$ and $539$ edges, respectively. 
The different graph measures are calculated, the first one is based on triangle 
motifs count and the second is based on local clustering.
The Watts-Strogatz clustering coefficient \cite{WS98} of a network of $N$ nodes is
\begin{equation}\label{Cws}
C^{W} = \frac1N \sum_i 2n_i / k_i(k_i-1) \ ,
\end{equation}
where $n_i$ denotes the number of direct links interconnecting the
$k_i$ nearest neighbors of node $i$.
An alternative is the ``global'' clustering coefficient \cite{globalC}
also called ``fraction of transitive triplets'',
\begin{equation}\label{Cg}
C^{\Delta} = \frac{\rm number \ of \ closed \ triplets}
{\rm number \ of \ connected \ triplets} \ .
\end{equation}

\begin{figure}[h]
\centering
\includegraphics[width=8cm]{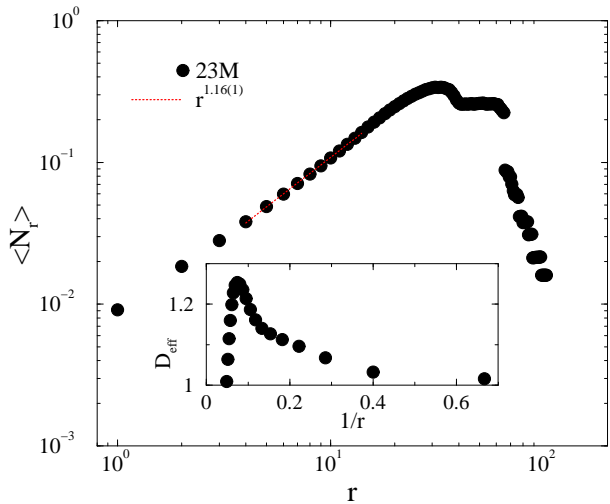}
\caption{\label{dim} Average number of nodes within topological distance $r$
in the $23$M graph. Dashed line shows a PL fit for $4<r<20$. 
Inset: local slopes defined in Eq.~(\ref{Deff}).}
\end{figure}

An important measure is the topological (graph) dimension $D$.
It is defined by
\begin{equation} \label{topD}
\langle N_r\rangle \sim r^D \ ,
\end{equation}
where $N_r$ is the number of node pairs that are at a topological
(also called ``chemical'') distance $r$ from each other 
(i.e.\ a signal must traverse at least $r$ edges to travel from one 
node to the other). The topological dimension characterizes how 
quickly the whole network can be accessed from any of its nodes:
the larger $D$, the more rapidly the number of $r$-th nearest
neighbors expands as $r$ increases.
To measure the dimension of the network we first computed the
distances from a seed node to all other nodes by running the
breadth-first search algorithm.
Iterating over every possible seed, we counted the number of nodes
$N_r$ with graph distance $r$ or less from the seeds and calculated
the averages over the trials in case of the largest, 23M network. 
As Fig.~\ref{dim} shows, an initial power law breaks down due to the 
finite network size. The small $\langle N_r\rangle$ values are due
to the sparsity and directedness of the graph.
We determined the dimension of the network, as defined by the scaling
law (\ref{topD}), by attempting a PL fit to the data 
$\langle N_r\rangle$ for the initial ascent. This suggests a slightly
super-linear behavior, increasing with the presence of central nodes.
To see the corrections to scaling we determined the effective exponents
of $D$ as the discretized, logarithmic derivative of (\ref{topD})
\begin{equation}  \label{Deff}
D_\mathrm{eff}(D+1/2) = \frac {\ln \langle N_r\rangle - \ln \langle N_{r+1}\rangle} {\ln(r) - \ln(r+1)} \ .
\end{equation}
These local slopes are shown in the inset of Fig.~\ref{dim} as the function
of $1/r$ and provide an increasing effective dimension due to the HV
nodes, before the finite size breakdown.
A similar analysis for the undirected US HV power grid with $N=4941$ 
nodes~\cite{USpg} results in $D > 2$. That means that this power grid has 
higher graph dimension than the embedding space due to some extra links. 
In our case the small number of HV links do not provide such contribution 
but the other, directed ones, which occur in the distribution sub-networks, 
dominate the whole topology.

\subsection{Comparison with other synthetic power grids}

The synthetic networks generated by the authors’ model is significantly different
to other synthetic networks, published in the literature. Such network generation
methods are introduced in \cite{Wang08,Schultz14,Patania15,Ma}.
The model proposed by \cite{Wang08} was created in order to model HV
transmission networks. The topology and the electrical parameters of the network
are created using specific random distribution functions, avoiding both
topological self-loops and islanded parts. The three-step process uses a
pre-determined number of nodes, with randomly distributed locations, selects
neighboring links of each bus and finally checks whether all nodes are connected.
The resulting networks have an average node degree between $2.66$ and $3.32$,
which is in range with real HV topologies with low node number.

Ma et al. \cite{Ma} presents “dual-stage constructed random graph”, generated by
an algorithm in two steps. First a random graph with one connected component is
created, then additional edges are added to the spanning tree. The algorithm is
tested on four networks; resulting average node degrees are between $2.42$ and $2.774$.

A random growth model is proposed by \cite{Schultz14} to create synthetic network 
topologies. A heuristic target function is used for redundancy and cost optimization 
during the initialization, and an attachment rule during the growth phase. 
The resulting networks have an average node degree of approximately $2.67$, and 
the degree distribution shows an exponential tail; both are characteristics of 
HV transmission networks. 
Schultz et al. write that “Despite this formally low level of topological connectedness, 
most links of a power grid are typically redundant minimum cost, redundancy”, which 
statement is true for high-voltage transmission networks, but not valid for distribution 
networks, which have a radial topology

A different synthetic network generation process is introduced in \cite{Patania15}, 
which connects nodes based on a local rule and is based on the epsilon-disk model. 
The three-step process consist of the assignment of nodal locations, types and attributes, 
a deterministic placement of the edges. The network generation method is tested on 
the Spanish power system, and resulting average node degrees are above $3$. 
Distribution of local clustering coefficients is also shown.

As it can be seen from the examples cited above, literature almost exclusively 
focuses on HV transmission networks when using synthetic network generation algorithms, 
creating undirected, unweighted and simple topologies with relatively low number of 
nodes, and average node degrees in the range of $2.4-2.8$. In comparison the network 
generation algorithm of the authors is able to create networks including HV 
transmission and MV and LV distribution parts as well. Such networks have significantly 
lower average node degrees due to the radial topology of distribution networks. 
Connectivity of the networks is also different, as the authors’ algorithm considers 
transformers of the substations as well (as an edges between two nodes, representing 
primary and secondary voltage levels). From a complex network analysis perspective, 
the generated graphs are undirected, but weighted, which is an important difference.

\section{Phase transition study}

We applied fourth order Runge-Kutta method (RK4 from Numerical Recipes)\cite{NumR}
to solve Eq.~(\ref{kur2eq}) on various networks. Step sizes:
$\Delta = 0.1, 0.01, 0.001$ as in~\cite{TL97} and the convergence 
criterion $\epsilon = 10^{-12}$ were used in the RK4 algorithm.
Generally the $\Delta =0.001$ precision did not improve the stability of the 
solutions except at large $K$-s, while $\Delta = 0.1$ was insufficient, 
so most of the results presented here are obtained using $\Delta = 0.01$.
The initial state was either fully synchronized: $\theta_i(t)=0$ or uniform
random distribution of phases: $\theta_i(t) \in (0,2\pi)$. 
We measured the Kuramoto order parameter:
\begin{equation}\label{ordp}
z(t_k) = r(t_k) \exp{i \theta(t_k)} = 1 / N \sum_j \exp{[i \theta_j(t_k)}] \ ,
\end{equation}
in a quenching process with a fixed $K$ by increasing the sampling time 
steps exponentially :
\begin{equation}
t_k = 1 + 1.08^{k} \ ,
\end{equation}
where $0 \le r(t_k) \le 1$ gauges the overall coherence and $\theta(t_k)$ is
the average phase. We solved (\ref{kur2eq}) numerically for $50$
independent initial conditions, with different $\omega_{i,0}$-s and 
determined the sample average: $R(t_k) = \langle r(t_k)\rangle$.
In the steady state, which occurred after $t > 100$, we measured
the standard deviation: $\sigma_R$ of $R(t_k)$ measured at $50$ 
sampling times. 

It is expected that for an infinitely large population of oscillators
the model exhibits a phase transition at some control parameter value $K$, 
separating a coherent steady state, with order parameter: 
$R(t\to\infty) > 0$ from an incoherent one $R(t\to\infty) = 0$ with
$1/\sqrt(N)$ finite size corrections.
\begin{figure}[h]
\centering
\includegraphics[width=8cm]{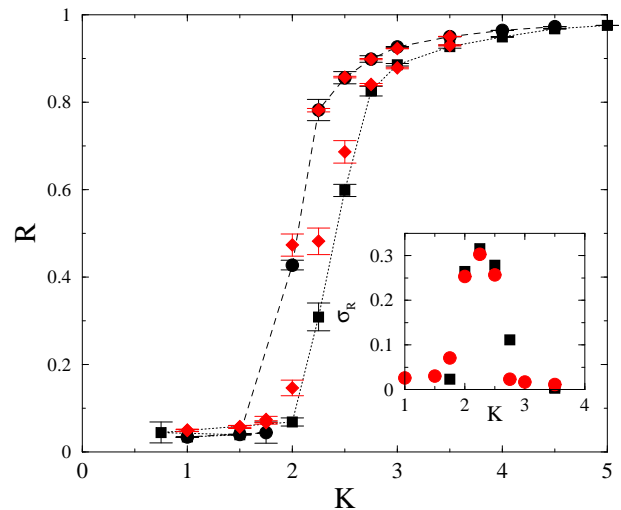}
\caption{\label{full-kur2G-ss} 
Hysteresis in the steady state order parameter in fully coupled networks
of sizes $N=1000$ (black boxes) and $N=500$ (red diamonds). 
Error bars show standard error of the mean. Inset: $\sigma_R(K)$
peaks for the two different network sizes investigated.}
\end{figure}
For the fully coupled network we recovered the first order transition,
known from the literature \cite{TL97}, as can be seen on 
Fig.\ref{full-kur2G-ss}. The synchronization transition occurs
around $K_c\simeq 2.25$, for $N=500$ and $N=1000$ both
and large hysteresis curves emerge as the consequence of 
different (fully ordered vs. randomized) initial conditions. At this
resolution only weak size dependence of the transition point is observable
in agreement with the results of \cite{29}.
The $\sigma_R(K)$ peak seems to be slightly higher in case of 
the larger lattice, as the inset of Fig.\ref{full-kur2G-ss} shows,
as opposed to the lower dimensional cases to be discussed later.
\begin{figure}[h]
\centering
\includegraphics[width=8cm]{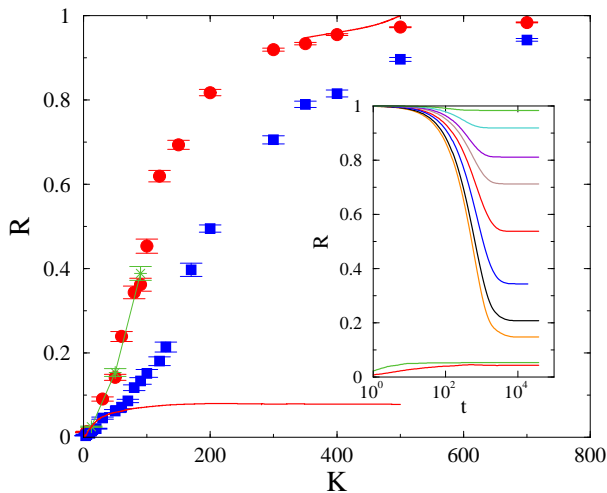}
\caption{\label{latt-kur2G-50-ss}
Phase synchronization transition in the steady state in 2D networks of
sizes $N=500 \times 500$ (red bullets), $L=1000 \times 1000$ (blue boxes) using $a=3$
and $N=500 \times 500$ (green stars ) using $a=1$. The red lines show the results 
using the adiabatic protocol, started from asynchronous (bottom) 
or synchronous (top) states in case of  $N=500 \times 500$ and $a=3$.
Error bars show standard deviation of the mean.
Inset: time dependence or $R(t)$, in case of the $N=500 \times 500$ lattice,
for control parameters: $K=700$, $350$, $200$, $150$, $100$, $80$, $60$, $50$
in case of synchronized initial condition (top to bottom curves) and for
$K=700$, $100$ de-synchronized initial condition (top to bottom curves).}
\end{figure}

In case of 2D lattices, with nearest neighbor interactions and
periodic boundary conditions, we found signatures of first order 
phase transitions with wide hysteresis loops  
(see Fig.\ref{latt-kur2G-50-ss}).
The synchronization emerges very slowly by increasing $K$. 
The finite size scaling study showed that the order parameter curves 
become smoother for larger $N$ and the transition point increases 
from $K \simeq 100$ ($L=500$) to $K \simeq 170$ ($L=1000$). 
Changing $a=3$ to $a=1$ did not cause visible differences.
The time dependence of the phase synchronization order parameter can 
also be seen on the inset of Fig.\ref{latt-kur2G-50-ss} for a lattice
of linear size $L=500$.
There are no signs of PL-s, instead the $R(t)$ curves converge quickly
to their steady state values at all $K$ values. 

To investigate the hysteresis in more detail we also applied an adiabatic 
procedure, in which following a start from the asynchronous state 
$K$ was increased gradually by $\Delta K = 0.02$ steps, separated by
$\Delta t_K = 1000$ intervals, containing $\Delta t_t = 900$ thermalization
and $\Delta t_m = 100$ measurement regions. In this protocol the 
measurements were done by linear  $\Delta t =1$ time-steps and 
averaging was performed over $48$ independent realizations of the 
quenched disorder.
As the lower (red) curve of Fig.\ref{latt-kur2G-50-ss} 
for $N=500 \times 500$ and $a=3$ shows, the synchronization remained 
very small up to $K=500$, in agreement with the steady state values 
of the quench with de-synchronized initial condition 
(see inset of Fig.\ref{latt-kur2G-50-ss}), but we could not reach 
the high branch of the solutions. 
When we started the adiabatic procedure from a synchronized initial condition: 
$K=500$, $R=1$ and decreased the coupling in the same way as in the up-sweep 
process we found agreement with the high branch of solutions, 
obtained with the quench procedure (see top red line vs red bullets 
of Fig.\ref{latt-kur2G-50-ss}).

The size of the hysteresis increased slightly by decreasing $a$ from $3$ 
to $1$, similarly as reported in \cite{29}. 
The former value was used in our subsequent, more detailed analyses 
in the hope of finding critical phase transitions as the consequence 
of network heterogenities.

\begin{figure}[h]
\centering
\includegraphics[width=8cm]{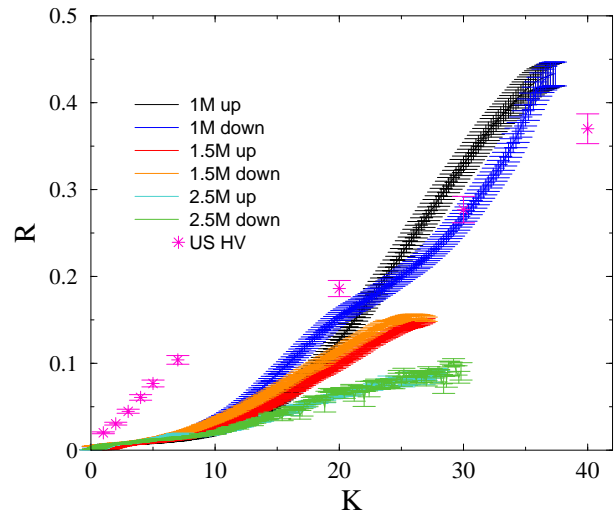}
\caption{\label{pow-lat-ss} 
Steady state order parameter for different power grid networks, using $a=3$
for 1M, 1.5M and 2.5M power grids (top to bottom curves), obtained by the
adiabatic protocol. Error bars show standard errors of the mean.
Pink stars correspond to the US HV power grid of size $N=4941$, for comparison.
We can see a vanishing synchronization and hysteresis by increasing the size.}
\end{figure}

However, we did not achieve this goal is case of the power grids we 
generated. Fig.~\ref{pow-lat-ss} shows that the transition in case of
our power grids is smooth, but a critical point with PL time dependencies 
could not been located.
Instead, fast relaxation to steady state values was observed using the
quench dynamics. 
The numerical solutions exhibited large fluctuations in the time dependencies 
and for large $K$-s the solutions become unstable, even with 
$\Delta=0.001$ precision.
Possible hysteresis curves now proved to be much narrower than in case 
of the 2D lattices. 
We have applied the adiabatic protocol, described in case of the
2D lattice, to provide more numerical evidence for this. Following
up-sweeps we turned back when reaching maxima at $K=37$ for 1M,   
at $K=27$ for 1.5M and at $K=30$ for 2.5M networks. 
The hysteresis curves look very narrow and in case of the 1M grid 
a looped "hysteresis" emerged, for all random realizations of the 
quenched disorder.
This strange behavior remained there even for $\Delta t_K = 2000$ intervals,
containing $\Delta t_t = 1900$ thermalization and $\Delta t_m = 100$ 
measurement regions. We suspect this the consequence of the
loopless topology of the 1M grid, different from the others.
In case of the 2.5M grid we could not see hysteresis within our
error margins (standard error of the mean).
So we find agreement with \cite{29} for Italian HV power grid, where  
"the transition is largely non-hysteretic, probably due to the low
value of the average connectivity in such a network."
  
Note, that without the weight normalization (\ref{wnorm})
the transition results would have appeared at much smaller
$K$ values if we had used the pure admittances as weights.
In case of the 1M grid we had the average: 
$\langle Y_{ij} \rangle = 854.13/\Omega$, 
while for the 2.5M network:  $\langle Y_{ij} \rangle = 763.05/\Omega$.
We have also considered the US HV network, in which case the 
results are similar to those of our synthetic networks.

\begin{figure}[h]
\centering
\includegraphics[width=8cm]{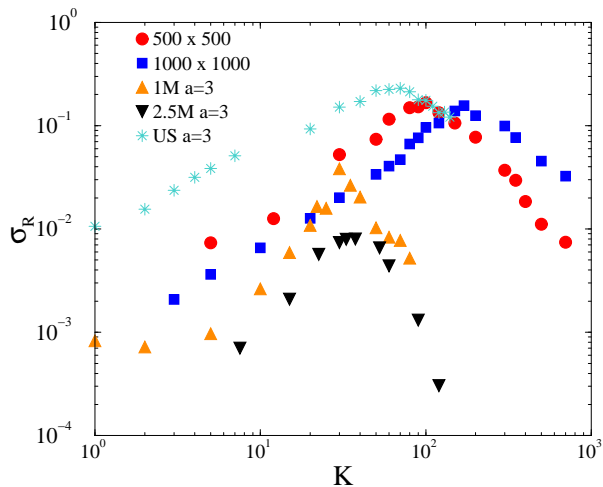}
\caption{\label{sigma} 
Fluctuation of the steady state order parameter for different networks 
with $a=3$.
Symbols: red bullets and blue boxes are for 2D lattices of linear sizes: 
$L=500, 1000$ respectively; up triangles are for 1M grids; 
down triangles are for 2.5M grids; stars correspond to the US HV grid.}
\end{figure}

Fig.~\ref{sigma} shows that the steady state order parameter
fluctuations ($\sigma_R$) remained bounded and the maxima of the
curves decreased when we increased the size of a given network system.
Thus, we don't see signatures of a singularity, a real phase 
transition in the thermodynamic limit, like in case of the Kuramoto 
model in low ($D<4$) dimensions \cite{HPClett}.
Fig.~\ref{sigma} also shows the results obtained for the
US HV power grid, containing $N=4941$ nodes, using $a=3$. 
On this small network the fluctuations are higher than those of the 
2D lattices and our power grid graphs.

\section{Power failure distributions}

\begin{figure}[h]
\centering
\includegraphics[width=8cm]{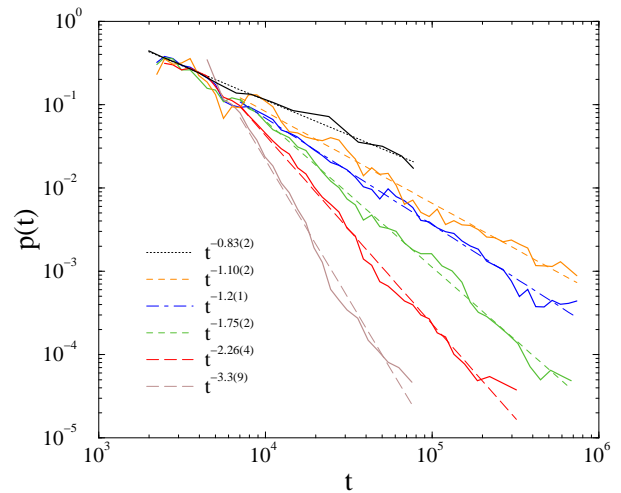}
\caption{\label{elo-av-latt-kur2GL-50}
Avalanche duration distribution for $1000\times 1000$ lattices 
for $a=3$ at different coupling values: $K=10$, $5$, 
$4$, $3$, $2$, $1$ (top to bottom solid curves). 
Dashed lines: PL fits for the distribution tails.}
\end{figure}

\begin{figure}[h]
\centering
\includegraphics[width=8cm]{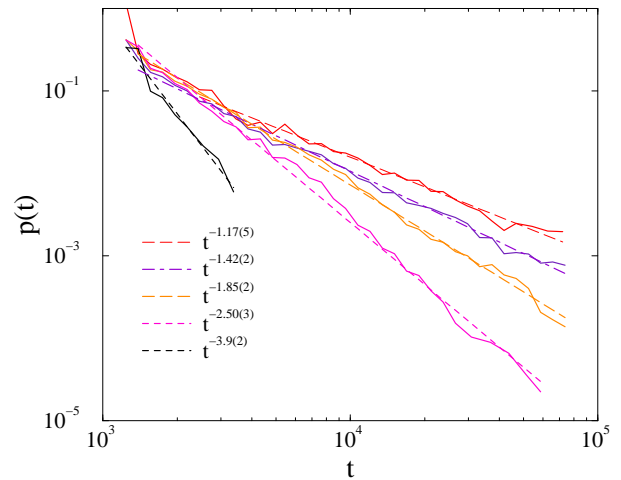}
\caption{\label{elo-Pow.eps}
Avalanche duration distribution in the 1M power grid  for $a=3$ and
different coupling values $K=0.7$, $0.6$, $0.5$, $0.4$, $0.2$
(top to bottom solid curves). Dashed lines: PL fits for the tails.}
\end{figure}

Power failure size dependence has been studied in different countries
and heavy tailed distributions were found, modelled by SOC models 
at the critical point of the their phase transitions \cite{Car,Car2}.
Since there is no real phase transition to synchronization in the 
second order Kuramoto model, we can investigate this issue in the
desynchronized state only. Following an electrical disturbance
local couplings can break down and the system is indeed 
in the non-synchronized state, where the effective $K$ is below 
the transition value of the finite system. Thus measuring the behavior
of the de-synchronization cascade can provide information about the 
seriousness of the power outage. 
We have investigated the avalanche duration distributions by starting 
the system from a fully synchronous state, quenching $K$ to a small
value and measuring the time until $R(t_k)$ fell below the 
threshold $R_T=1/\sqrt{N}$, related to the order parameter value 
of the incoherent phase. 
In this measurement we averaged over $\simeq 10^4$ runs, using independent
random $\omega_{i,0}$ intrinsic frequencies.
As we can see on Figs.~\ref{elo-av-latt-kur2GL-50},\ref{elo-Pow.eps} 
in the incoherent phase $K$-dependent PL decay tails emerge, reminiscent to
GPs in other heterogeneous network models \cite{CCdyncikk},
very differently from an exponential decay of a random system. 
Even with this large sample number the results exhibit oscillations, especially
approaching the transition region, where reaching $R_T$ requires long times.
Thus we limited the range of $K$-s, where the decay was faster than linear.
The range of the PL region can be estimated by the $K$ values, where linear 
behavior can be fitted on the $p(t)$ tails. This provides $K < \simeq 5$
for the 2D lattice with $N=10^6$, $K < \simeq 1$ for the 1M power grid 
and $K < \simeq 7$ for the US HV network. 
In the latter cases the PL region is enhanced by the quenched topological
heterogeneity.
In case of a 2D lattice, without any quenched disorder, i.e. $\omega_{i,0}=0$,
but with an additive, annealed Gaussian frequency noise of unit variance
in (\ref{kur2eq}) we could not find PL tails, but fast decaying $p(t)$ 
distributions only.

\section{Further extensions}

Recently, it has been shown that a large number of decentralized 
generators, rather than a small number of large power plants, provide enhanced
synchronization together with greater robustness against structural failures
\cite{Rhoden12,Witthaut12,Rhoden14,Lee-Kim}. 
Here we studied effects of additional time-dependent stochastic noise
to Eq.~(\ref{kur2eq}).
We added the same, time dependent random variable to $\omega_{i,0}$ 
following the probability distribution
\begin{equation}
p(\omega) \sim \pm e^{-0.06\omega} \ ,
\end{equation}
which is similar to what can be read-off from the MAVIR 
frequency fluctuation data \cite{MAVIR}.

Another attempt was the addition of a space and time independent, 
uncorrelated Gaussian noise with $\sigma=3$ variance, describing a 
stochastic Kuramoto model.
Neither of these modifications gave relevant changes in the dynamical behavior.
The annealed noise decreased the order parameter as well as its fluctuations slightly.

We have also performed preliminary calculations for bimodal Gaussian 
$\omega_{i,0}$ distributions, modelling a coupled consumer/motor system 
\cite{29}. Following the initial, large fluctuations the order parameter relaxes
in a similar way as before, but to smaller synchronization values. 
More detailed study of this scenario will be published later.

\section{Conclusions}

We compared the phase synchronization of the second order Kuramoto model on
fully coupled, 2D lattices and real power grid networks. For this purpose
we generated large synthetic networks in order to extrapolate to infinite sizes, 
with characteristics or real power grids. These contain millions of nodes 
and directed, weighted edges. Our networks exhibit hierarchical modular 
structure, low clustering and topological dimensions. 

Real phase transition could be observed on the fully coupled graph, showing
hysteresis and first order transition. On lower graph dimensional systems, 
like in the power grids or in 2D lattices smooth crossover occurs at
higher global coupling values.
The transition peak locations, obtained by the maximum of the fluctuations 
of $R$ are lower for the power grids: $K \simeq 20 - 30$,
than in case of the 2D lattices: $K \simeq 100 - 170$ of similar sizes. 
The magnitudes of the fluctuations are also lower on the power grids 
than in the corresponding 2D lattices, albeit a decreasing tendency can be 
found by increasing the inertia.

The addition of a stochastic noise to Eq.~(\ref{kur2eq}), modeling random 
frequencies of distributed energy sources does not affect the 
synchronization too much. Even a strong Gaussian noise with $\sigma=3$ 
variance decreases the order parameter by $20\%$ few percent at most.
These results point out better electrical performances in the
heterogeneous networks than what simple homogeneous approximations
could predict.

Scale-free tails of the avalanche duration can be observed
below the transition point with $K$-dependent slopes.
The size of this scale-free region increases with the amount
of quenched disorder. For pure 2D lattices we could not
found PL tails, but quick decays only. 
This is similar to the Griffiths effects, which can occur in 
disordered phases of magnets in the presence of slowly decaying, 
rare, but large ordered regions.
However, in the lack of a real critical phase transition in the 
continuum limit we cannot call this a Griffiths phase.
Probably our results are related to the "frustrated synchronization"
phenomena, reported recently in case of the Kuramoto model, 
where modules, as rare regions, synchronize to different phases 
\cite{Frus,FrusB}.
Understanding rare region effects in more detail in power grid 
models should be a subject of further studies.

We emphasize that mechanism that would create self-organized 
criticality has not been assumed in our model, still we see PL 
tails of event duration with similar exponents as those of
the reported blackout sizes in various electrical failure data 
\cite{Car2}. 
It is an open question how such additional, competing forces would
modify our results.

\section{Acknowledgments}

We thank R\'obert Juh\'asz and S.C. Ferreira for the useful discussions
and comments. Support from the MTA-EK special grant and the
Hungarian research fund OTKA (K109577) is acknowledged.
The VEKOP-2.3.2-16-2016-00011 grant is supported by the European Structural
and Investment Funds jointly financed by the European Commission and the
Hungarian Government.
Most of the numerical work was done on NIIF supercomputers of Hungary.


\begin{thebibliography}{50}

\bibitem{1} Andersson A. et al., Causes of the 2003 major grid blackouts
in North America and Europe, and recommended means to improve system dynamic
performance. {\it IEEE Trans. Power Syst.} {\bf 20,} 1922–1928 (2005).
\bibitem{Ace05} Acebr\'on J. A., Bonilla L. L., P\'erez Vicente C. J., Ritort F., Spigler R.,
The Kuramoto model: A simple paradigm for synchronization phenomena.
{\it Rev. Mod. Phys.} {\bf 77,} 137 (2005).
\bibitem{2} Arenas A., Diaz-Guilera A., Kurths J., Moreno Y., and Zhou C. S.,
Synchronization in complex networks.
{\it Phys. Rep.} {\bf 469,} 93–153 (2008).
\bibitem{fila} Filatrella G., Nielsen A. H. and Pedersen N. F.,
Analysis of a power grid using a Kuramoto-like model.
{\it Eur. Phys. J. B} {\bf 61,} 485–491 (2008).
\bibitem{3} Carareto R., Baptista M. S. and Grebogi C.,
Natural synchronization in power grids with anti-correlated units.
{\it Commun. Nonlinear Sci. Numer. Simul.} {\bf 18,} 1035–1046 (2013).
\bibitem{4} Choi Y.-P., Ha S.-Y. and Yun S.-B.,
Complete synchronization of Kuramoto oscillators with finite inertia.
{\it Physica D} {\bf 240,} 32–44 (2011).
\bibitem{5} Choi Y.-P., Li Z., Ha S.-Y., Xue X. and Yun S.-B.,
Complete entrainment of Kuramoto oscillators with inertia on networks via
gradient-like flow.
{\it J. Differ. Equations} {\bf 257,} 2591–2621 (2014).
\bibitem{12} Dorfler F. and Bullo F., Synchronization and transient
stability in power networks and non-uniform kuramoto oscillators.
{\it SIAM J. Control Optim.} {\bf 50,} 1616 (2010).
\bibitem{13} Dorfler F. and Bullo F., Synchronization in complex
networks of phase oscillators: A survey.
{\it Automatica} {\bf 50,} 1539–1564 (2014).
\bibitem{18} Fortuna L, Frasca M. and Fiore  A. S.,
Analysis of the Italian power grid based on kuramoto-like model.
{\it Proceedings of Physcon 2011}, (Leon, Spain, 5–8).
\bibitem{29} Olmi S., Navas A., Boccaletti S., and Torcini A.,
Hysteretic transitions in the Kuramoto model with inertia.
{\it Phys. Rev. E} {\bf 90,} 042905 (2014).
\bibitem{32} Pinto R. S. and Saa A., Synchrony-optimized networks
of Kuramoto oscillators with inertia. {\it Physica A} {\bf 463,} 77–87 (2016).
\bibitem{36} Schmietendorf K., Peinke J., Friedrich R. and Kamps O.,
Self-organized synchronization and voltage stability in networks of synchronous machines.
{\it Eur. Phys. J. Spec. Top.} {\bf 223,} 2577–2592 (2014).
\bibitem{Gryz} Grzybowski J. M, Macau E. E. and Yoneyama T.,
On synchronization in power grids modelled as networks of second-order Kuramoto oscillators.
{\it Chaos.} {\bf 26,} 113113 (2016).
\bibitem{kura} Y. Kuramoto, {\it Chemical Oscillations, Waves, and Turbulence},
(Springer, Berlin, 1984).
\bibitem{TL97} Tanaka H.-A., Lichtenberg A. J. and Oishi S., First order phase transition
resulting from finite inertia in coupled oscillator systems. 
Phys. Rev. Lett. {\bf 78,} 2104–2107 (1997).
\bibitem {odorbook} \'Odor G., {\it Nonequilibrium Lattice Systems},
(World Scientific, 2008 Singapore).
\bibitem{Car} Carreras B. A., Newman D. E., Dobson I., Poole A. B., Evidence for self-organized
criticality in a time series of electric power system blackouts.
{\it IEEE Transactions on Circuits and Systems I: Regular Papers} {\bf 51,} 1733-1740 (2004).
\bibitem{round} Martin P. V., Bonachela J. A. and Mu\~noz  M. A.,
Quenched disorder forbids discontinuous transitions in nonequilibrium 
low-dimensional systems. {\it Phys. Rev. E} {\bf 89,} 012145 (2014).
\bibitem{Bak} Bak P., Tang C. and Wiesenfeld K., Self-organized criticality.
{\it Phys. Rev. A} {\bf 38,} 364~V374 (1988).
\bibitem{Vojta} Vojta T., Rare region effects at classical, quantum and nonequilibrium
phase transitions {\it J. Physics A: Math. and Gen.} {\bf 39,} R143-R205 (2006).
\bibitem{Griffiths} Griffiths R. B., Nonanalytic Behavior Above the Critical Point in a
Random Ising Ferromagnet.
{\it Phys. Rev. Lett.} {\bf 23,} 17-19 (1969).
\bibitem{burstcikk} \'Odor, G. Slow, bursty dynamics as a consequence of quenched
network topologies.
{\it Phys. Rev. E} {\bf 89,} 042102 (2014).
\bibitem{Pagani} Pagani G. A., {\it From the Grid to the Smart Grid, Topologically},
PhD dissertation, (Rijskuniversiteit Groningen, 2014).
\bibitem{Gomez} G\'omez T., Mateo C., S\'anchez \'A., Frias P., Cossent R., 
Reference Network Models: a Computational Tool for Planning and Designing Large-Scale Smart 
Electricity Distribution Grids in Khaitan S. K. and Gupta A. (Eds.) {\it HPC in power and Energy Systems}, 
247-279. (Springer Science \& Business Media, 2013)
\bibitem{Ma} Ma S., Yu Y., Zhao L., Dual-stage constructed random graph algorithm to generate 
random graphs featuring the same topological characteristics with power grids. 
{\it J. Mod. Power Syst. Clean Energy} {\bf 5,} 683-695 (2017).
\bibitem{Pahwa} Pahwa S., Scoglio C., Scala A., Abruptness of Cascade Failures in Power Grids. 
{\it Scientific Reports} {\bf 4,} 3694 (2014).
\bibitem{Wang} Whang Z., Thomas R. J., Random Topology Power Grid Modeling and 
Automated Simulation Platform. {\it CERTS Review} {\bf 5-6,} (2014).
\bibitem{USpg} US power grid,
\url{http://konect.uni-koblenz.de/networks/opsahl-powergrid}
\bibitem{Wang08} Wang Z., Thomas R. J, Scaglione A.,
Generating Random Topology Power Grids.
{\it  Proc. 41st Hawaii International Conference on Science Systems}.
\bibitem{Schultz14} Schulz P., Heitzig J. and Kurths J.,
A random growth model for power grids and other spatially embedded infrastructure networks.
{\it Eur. Phys. J. Special Topics} {\bf 223,} 2593–2610 (2014).
\bibitem{Patania15} Patania A, et al.,
Complex Systems Techniques applied to Power Transmission Expansion Planning. 
Part I : Generating Random Networks that are Consistent with Power Transmission.
\bibitem{NumR} Numerical Recipes, \url{http://numerical.recipes}
\bibitem{HPClett} Hong H, Park H. and Choi M. Y., Collective synchronization in 
spatially extended systems of coupled oscillators with random frequencies.
{\it Phys. Rev. Lett.} {\bf 72,} 036217 (2005).
\bibitem{CCdyncikk} \'Odor G., Critical dynamics on a large human Open Connectome network.
{\it Phys. Rev. E} {\bf 94,} 062411 (2016).
\bibitem{WS98} Watts D. J. and Strogatz S. H., Collective dynamics of “small-world” networks.
{\it Nature} {\bf 393,}  440--442 (1998).
\bibitem{globalC} Newman M. E., Moore C. and Watts  D. J., Mean-field solution of the small-world
network model. {\it Phys. Rev. Lett.} {\bf 84,} 3201–-320 (2000).
\bibitem{Rhoden12} Rohden M., Sorge A., Timme M. and Witthaut D.,
Self-Organized Synchronization in Decentralized Power Grids.
{\it Phys. Rev. Lett.} {\bf 109,} 064101 (2012).
\bibitem{Witthaut12} Witthaut TD. and Timme M., Braess's paradox in oscillator networks, 
desynchronization and power outage.
{\it New J. Phys.} {\bf 14,} 083036 (2012).
\bibitem{Rhoden14} Rohden M., Sorge A., Witthaut D and Timme M., 
Impact of network topology on synchrony of oscillatory power grids
{\it Chaos} {\bf 24,} 013123 (2014).
\bibitem{Lee-Kim} Lee M. J. Lee and Kim B. J., Spatial uniformity in the power grid system,
{\it Phys. Rev. E} {\bf 95,} 042316 (2017).
\bibitem{MAVIR} {\it Data of the Hungarian electrical system},
\url{https://www.mavir.hu/documents/10258/45985073/MAVIR_VER_2017_web.pdf}, (MAVIR 2016)
\bibitem{Frus} Villegas P., Moretti P. and Mu\~noz M. A., Frustrated hierarchical synchronization
and emergent complexity in the human connectome network.
{\it Scientific Reports} {\bf 4,} 5990 (2014).
\bibitem{FrusB} Mill\'an A. P., Torres J. J. and Bianconi B., Complex network geometry and
frustrated synchronization, {\it arXive:1802.00297}.  
\bibitem{Car2} Dobson I., Carreras B. A., Lynch V. E. and Newman D. E.,
Complex systems analysis of series of blackouts: Cascading failure, critical points
and self-organization. {\it Chaos} {\bf 17,} 026103 (2007).
\end{thebibliography}
\end{document}